\documentstyle[manuscript,aps]{revtex}
\begin{document}
\draft
\title{Einsteinian Strengths and Dynamical Degrees of Freedom for Alternative
Gravity Theories}
\author{Janusz Garecki}
\address{Institute of Physics, University of Szczecin, Wielkopolska 15; 70-451
Szczecin, POLAND}
\date{\today}
\maketitle
\begin{abstract}
In the paper we present the results of the our calculations of the Einsteinian
strengths $S_E(d)$ and numbers dynamical degrees of freedom $N_{DF}(d)$ for
alternative gravity theories in $d\geq 4$ dimensions. At the first part we
consider the numbers $S_E(d)$ and $N_{DF}(d)$ for metric-compatible and
quadratic in curvature (or quadratic in curvature and in torsion) gravity. We show that in the all set of the metric-compatible
quadratic gravity in $d\geq 4$ dimensions the 2-nd order {\it Einstein-Gauss-Bonnet}
theory has {\it the smallest numbers} $S_E(d)$ and $N_{DF}(d)$, i.e., this
quadratic theory of gravity has {\it the strongest field equations}. From the
physical point of view this theory is the best one quadratic and metric-compatible  theory
of gravity in $d\geq 4$ dimensions.

At the second part of the paper we study the numbers $S_E(d)$ and $N_{DF}(d)$
in $d\geq 4$ dimensions for some other alternative gravity theories
which are popular recently. 

We finish our paper with some conclusions.
\end{abstract}
\newpage 
\section{Introduction}
 The notion of ``strength of the field equations'' is a concept which enables us to compare different
systems of the field equations. It was introduced by Einstein [1] in
order to analyze systems of partial differential equations for physical
fields. Later this notion was examined and effectively used in field
theory by several authors [2--7]. In particular B.F. Schutz [3] pointed
out that the Einsteinian strength of the field equations is closely
connected with the number of the dynamical degrees of freedom which
these equations admit in Cauchy problem.

The idea of ``strength'' is the following (see e.g. [2]). Suppose we have an
analytic field function $\Phi$ of $d$ variables. We can expand it in Taylor series
and the totality of its coefficients describe the field completely. Let us
consider the terms of the nth-order of differentiation in the Taylor
expansion; these are of the form
\begin{equation}
\partial_{i_1}\partial_{i_2}...\partial_{i_n}\Phi, ~~~~i_j = 1,2,...,d,
\end{equation}
and the total number of such coefficients is 
\begin{equation}
N_n(d) = {d\brack n} := {d+n-1\choose n} = {(d+n-1)!\over(d-1)! n!}.
\end{equation}
If the function $\Phi$ satisfies some field equations or constraints,
these give several relations $M_n(d)$ between the nth-order
coefficients, thereby reducing the number of coefficients left free to
be assigned arbitrary values. Let us denote the number of coefficients
left free by $Z_n(d)$. By definition we have 
\begin{equation}
Z_n(d) = N_n(d) - M_n(d).
\end{equation}
One can prove (see e.g. [2]) that
\begin{equation}
Z_n(d) = N_n(d) - M_n(d) = {d\brack n}\bigl(z_0 + {z_1\over n} +
{z_2\over n^2} + ...\bigr).
\end{equation}
$z_0$ gives the number of functions of $d$ variables $(x^1,x^2,...,
x^n)$ left free. For the {\it absolutely compatible} systems $z_0 =
0$.\footnote{$z_0 = 0$ for any physically reasonable system (see e.g. [1,2,7])} $Z_n(d)$
is always $\geq 0$ for all $n$. 

The coefficient of ${1\over n}$, $z_1 =: S_E(d)$ give measure of the
Einsteinian {\it strength} of the system of field equations under
consideration. As one can easily see the Einsteinian strength $S_E(d)$
can also be defined as the coefficient of $1/n$ in the ratio
${Z_n(d)\over{d\brack n}}$. From practical reasons we will use in the
following this last definition of $S_E(d)$.

Larger the value of $S_E(d)$, weaker is the system. Of course, in the
field theory {\it we need the strongest systems}.

The limit for large $n$ of
\begin{equation}
{Z_n(d)\over{d-1\brack n}}\asymp
{S_E(d)\over(d-1)} =: N_{DF}(d)
\end{equation}
is the number of free functions of $(d-1)$ variables in the theory [3--7] necessary to determine a
solution locally. For hyperbolic systems this is the amount of Cauchy data,
i.e., the number $N_{DF}(d)$ determines {\it the amount of dynamical freedom in
the system}.

The symbol $\asymp$ means equality in the highest powers of $n$.
$n\rightarrow\infty$.  

The paper is organized as follows. In Chapter II we discuss the number
$S_E(d)$ and $N_{DF}(d)$ for metric-compatible quadratic theories of
gravity. We will divide these theories onto the following two classes:
\begin{enumerate}
\item The purely metric quadratic in curvature theories of gravity ({\bf
PMQG}) [8--15], and 
\item The metric-compatible quadratic in curvature (or quadratic in
curvature and torsion) gravity theories with torsion ({\bf PGT}) [16--22]).
\end{enumerate}

The {\bf PMQG} theories give us a geometrization of the improved, symmetric energy-momentum tensor $T^{ik} = T^{ki}$ for
matter in the framework of the Riemann geometry. The gravitational
equations are here obtained by use {\it Hilbert variational principle} and
have the following general form
\begin{equation}
{\delta(\sqrt{\vert g\vert}L_g)\over\delta g^{ab}} = {\delta(\sqrt{\vert
g\vert}L_{mat})\over\delta g^{ab}} \bigl(=1/2\sqrt{\vert
g\vert}T_{ab}\bigr).
\end{equation}
Here we have 10 field equations of the 4th-order in general. In
consequence we have here problems with ghosts and tachyons in
weak field approximation, and with Newtonian limit.

The metric-compatible quadratic theories of gravity with torsion called
``Poincar\'e gauge quadratic field theories of gravity'' ({\bf PGT})
give us a geometrization of the canonical pair $(_c T^{ik}\not= _c
T^{ki}, ~_c S^{ikl} = (-) _c S^{kil})$ of matter tensors in the
framework of the more general Riemann-Cartan geometry. The gravitational
field equations are here obtained by use the {\it Palatini variational
principle}. 

In the Palatini variational principle we take $(g_{ik};~\Gamma^i_{~kl})$
as independent geometrical variables or, equivalently, an orthonormal
tetrads field $h^{(a)}_{~~~i}(x)$ and {\it Lorentz connection} ($\equiv$
spin connection) $\Gamma_{(a)(b)}^{~~~~~~l}$, where the indices inside
round brackets mean tetrads indices. This variational principle
leads us to Einstein-Cartan-Sciama-Kibble ({\bf ECSK}) theory of gravity
for linear gravitational Lagrangian $L_g = \chi R =: L_E$ and to its
generalization -- {\bf PGT} for a quadratic in curvature (and in
torsion) gravitational Lagrangian.
Here we have 40 field equations of 2nd-order w.r.t. $h^{(a)}_{~~~k}$ and
$\Gamma^{(i)(k)}_{~~~~~~l}$ or, equivalently, 40 equations of 3rd-order
w.r.t. metric and contortion $K^i_{~kl}$ where contortion is defined by
the decomposition 
\begin{equation}
\Gamma^i_{~kl} = {i\brace kl} + K^i_{~kl}.
\end{equation}

In this case we also have problems with ghosts and tachyons in weak field
approximation, and with Newtonian limit.

The gravitational field equations have the form (if we take an
orthonormal tetrads field $h^{(a)}_{~~~k}(x)$ and Lorentz connection
$\Gamma_{(i)(k)}^{~~~~~~l}$ as independent geometrical variables)
\begin{eqnarray}
{\delta(\sqrt{\vert g\vert}L_g)\over\delta h^{(a)}_{~~~k}}& = &
{\delta(\sqrt{\vert g\vert}L_{mat})\over\delta h^{(a)}_{~~~k}} 
~\bigl(= \sqrt{\vert g\vert} _c T_{(a)}^{~~~k}\bigr)\nonumber \\
{\delta(\sqrt{\vert g\vert}L_g)\over\delta\Gamma^{(i)(k)}_{~~~~~~l}} &=&
{\delta(\sqrt{\vert g\vert}L_{mat})\over\delta\Gamma^{(i)(k)}_{~~~~~~l}}
~\bigl( = \sqrt{\vert g\vert} _c S_{(i)(k)}^{~~~~~~l}\bigr),
\end{eqnarray} 
plus {\it metricity constraints}
\begin{equation}
D g_{kl} = 0,
\end{equation}
where $D$ means the exterior covariant derivative.

We have
\begin{eqnarray}
_c T_i^{~k}& =& h^{(a)}_{~~~i}{} _c T_{(a)}^{ ~~~k} \nonumber \\
_c S_{ik}^{~~l}& =& h^{(a)}_{~~~i}{} h^{(b)}_{~~~k}{}~ _c
S_{(a)(b)}^{~~~~~~l}.
\end{eqnarray}
In Chapter III we analyze $S_E(d)$ and $N_{DF}(d)$ for other
alternative gravity theories, which are popular recently. Namely, we
discuss these numbers for:
\begin{enumerate}
\item Teleparallel equivalent of general relativity ({\bf TEGR}) [23--25],
\item Teleparallel {\it new general relativity} ({\bf TNGR}) of Hayashi
and Shirafuji [26--27],
\item General relativity + scalar field ($\equiv$ scalar-tensor theories
of gravity), (see e.g. [28--29]), and
\item The most general {\it metric-affine} gauge theory of gravity ({\bf
MAG}) developed by F.W. Hehl et al., (see e.g. [30--33]).
\end{enumerate}

We would like to remark that the gravitational Lagrangians for {\bf
TEGR} and {\bf TNGR} are quadratic in metric-compatible torsion, and the
gravitational Lagrangian $L_g$ for {\bf MAG} is quadratic in irreducible
parts of curvature, nonmetricity and torsion.

We finish our paper with some Conclusions. Our main conclusions are the
following: from the point of view of the considered numbers $S_E(d)$ and
$N_{DF}(d)$, $d\geq 4$, 
\begin{enumerate}
\item The 2nd-order {\it Einstein-Gauss-Bonnet} theory of gravity is the
best one quadratic and metric-compatible gravity theory in $d\geq 4$
dimensions, 
\item {\bf MAG} has to week field equations (and too much free parameters
 -- 28) in order to be a reasonable alternative theory of gravity.
\end{enumerate}
\section{Strengths and Dynamical Degrees of Freedom for Quadratic
Gravity Theories in $d\geq 4$ Dimensions}
Here we will present the results of calculations of the Einsteinian
strengths $S_E(d),~d\geq 4$, and numbers of dynamical degrees of freedom
$N_{DF}(d), ~d\geq 4$ for a typical {\bf PMQG} (4th-order in general) which follows from the
Lagrangian 
\begin{equation}
L_g = \chi R + c_0 R^2 + c_1\vert Ric\vert^2 + c_2\vert Riem\vert^2,
\end{equation}
where $\chi,~c_0,~c_1,~c_2$ are some dimensional constants, and for a
typical {\bf PGT} with torsion.

The gravitational Lagrangian $L_g$ for {\bf PGT} can only be quadratic
in curvature like (11), but admitting torsion, or can contains terms
quadratic in curvature like (11) plus terms quadratic in irreducible
components of torsion. The most general Lagrangian $L_g$ for {\bf PGT}
was given, e.g., by Hayashi and Shirafuji [17].

We have the following $\\$ 
\leftline{\it Proposition 1}

The number $Z_n(d)$ of the free coefficients of order $n$ in Taylor
expansion of an analytic solution to the vacuum field equations which
follow from (11) is equal
\begin{equation}
Z_n(d)\asymp {d\brack n} {12{d\choose 3}\over n}\asymp 2d(d-2){d-1\brack
n},
\end{equation}
where
\begin{equation}
{d\brack n} := {n+d-1\choose n} = {(n+d-1)!\over n!(d-1)!},
\end{equation}
and the symbol $\asymp$ means equality in the highest powers of $n$.

$Proof \\$
Since we are working with purely metric theory of gravity then we have
${d(d+1)\over 2} = {d+1\choose 2}$ unknown metric functions which
determine an analytic solution. ${d+1\choose 2}$ functions of $d$
variables ($d$ coordinates $x^0,x^1,...,x^{d-1}$) give the total number
${d+1\choose 2}{d\brack n}$ of the nth-order coefficients. But our
theory is {\it generally covariant}. This means that the action integral
\begin{equation}
S = \int\limits_{\Omega}\sqrt{\vert g\vert}L_gd\Omega
\end{equation}
and the vacuum field equations 
\begin{equation}
L_{ab} := {\delta(\sqrt{\vert g\vert}L_g)\over\delta g^{ab}} = 0
\end{equation}
are invariant under the grupoid $Diff~M_d$ of coordinate
transformations.

The invariance of the action integral leads us to $d$ {\it generalized
Bianchi identities} of the form
\begin{equation}
\nabla^a{} L_{ab}\equiv 0.
\end{equation}

The invariance of the field equations means that if $g_{ij}(x)$ are
representatives of the components of the metric tensor then so are the
other representatives related to $g_{ij}(x)$ by a coordinate
transformation
\begin{equation}
g_{i^{\prime}j^{\prime}}(x^{\prime})= {\partial x^i\over\partial
x^{i^{\prime}}}{\partial x^j\over\partial x^{j^{\prime}}}g_{ij}(x).
\end{equation}
The last equation tell us that from the number of the nth-order
arbitrary coefficients of $g_{ij}$ one should subtract $d{d\brack n+1}$,
i.e., gauge freedom establishes $d{d\brack n+1}$ coefficients.

Let us now consider the number of relations between the nth-order
coefficients due to the vacuum field equations (15). These field equations
form the system ${d(d+1)\over 2} = {d+1\choose 2}$ equations of the
4th-order (in general). These therefore give ${d+1\choose 2}{d\brack
n-4}$ relations between the nth-order coefficients of $g_{ij}$. But not
all these relations are independent because the field equations (15)
satisfy $d$ {\it differential identities} (16) which are of the
5th-order. So, only 
\begin{equation}
{d+1\choose 2}{d\brack n-4} - d{d\brack n-5}
\end{equation}
relations between nth-order coefficients of $g_{ij}$ are independent.

Summing up, we have the following number $Z_n(d)$ of the free
coefficients of order $n$ in Taylor's expansion of an analytic solution
$g_{ij}(x)$ of the vacuum equations (15)
\begin{eqnarray}
Z_n(d) &=& {d+1\choose 2}{d\brack n} - d{d\brack n+1}\nonumber \\
&-&\Bigl\{{d+1\choose 2}{d\brack n-4} - d{d\brack n-5}\Bigr\}.
\end{eqnarray}
By use the asymptotic formulas
\begin{equation}
{d\brack n-k} = {d\brack n}\Bigl\{1 - {k(d-1)\over n} + 0({1\over
n^2})\Bigr\},
\end{equation}
\begin{equation}
{d\brack n} = {n\over(d-1)}{d-1\brack n}\bigl\{1 + 0({1\over n})\bigr\},
\end{equation}
one can easily obtain from (19)
\begin{eqnarray}
Z_n(d)&\asymp& {d\brack n}{12{d\choose 3}\over n}\nonumber \\
&\asymp& 2d(d-2){d-1\brack n}.
\end{eqnarray}
{\bf QED}

One can read from (22) that the Einsteinian strength $S_E(d)$ for the
quadratic gravity with gravitational Lagrangian (11) reads 
\begin{equation}
S_E(d) = 12 {d\choose 3},
\end{equation}
while the number dynamical degrees of freedom 
\begin{equation}
N_{DF}(d) = {S_E(d)\over(d-1)} = 2d(d-2).
\end{equation}
Iff $d=4$ then we have from the two last formulas
\begin{equation}
S_E(4) = 48, ~~N_{DF}(4) = 16.
\end{equation}

In the special cases the numbers $S_E(d)$ and $N_{DF}(d)$ can be
smaller. For example, in the case $3c_0 + c_1 = 0$ ($\equiv$
Bach-Weyl-Einstein theory) we have
\begin{eqnarray}
Z_n(d)&\asymp& {d\brack n}{2(d-1)[(d-1)^2 - 2]\over n}\nonumber \\
&\asymp& {d-1\brack n}2[(d-1)^2 -2],
\end{eqnarray}
i.e., here we have 
\begin{equation}
S_E(d) = 2(d-1)[(d-1)^2 -2], ~~N_{DF}(d) = 2[(d-1)^2 -2].
\end{equation}
Iff $d=4$, then we have for the Bach-Weyl-Einstein theory
\begin{equation}
S_E(4) = 42, ~~ N_{DF}(4) = 14.
\end{equation}

The other special case is given by $c_1 = c_2 = 0$, i.e., by $L_g = \chi
R + c_0 R^2$. In this case we have 
\begin{equation}
S_E(d) = (d-1)^2(d-2) , ~~N_{DF}(d) = (d-1)(d-2),
\end{equation}
i.e., we have in this case the same values of the numbers $S_E(d)$ and
$N_{DF}(d)$ as in the case of the so-called {\it scalar-tensor theories
of gravity} with linear gravitational Lagrangian $L_g$ (See Section
III.D).

Iff $d=4$ , then we have in the last case
\begin{equation}
S_E(4)  = 18, ~~ N_{DF}(4) = 6.
\end{equation}
These results are not surprising because the higher-order gravity theories
with $L_g = f(R), ~~f^{\prime}(R)\not = 0$ are dynamically equivalent
general relativity ({\bf GR}) plus a new scalar field $\Psi$ (see e.g
[15]).

We would like to emphasize that there exist an interesting example of
the {\bf PMQG} called {\it Einstein--Gauss--Bonnet theory} ({\bf EGBT})
[34--41] which has gravitational Lagrangian $L_g$ of the form 
\begin{eqnarray}
L_g &=& L_E + L_{GB} = L_E \nonumber \\
&+& \alpha\bigl(R^{iklm} R_{iklm} - 4R^{ik} R_{ik} + R^2\bigr),
\end{eqnarray}
where $\alpha$ is a new coupling constant.

The Lagrangian $L_{GB}$ is  called {\it Gauss--Bonnet} or {\it Lovelock}
Lagrangian. 

The field equations  of the {\bf EGBT} are of the 2nd-order for $d\geq
4$ (iff $d=4$, then these field equations are simply Einstein equations) and this quadratic theory of gravity {\it admits no
ghosts or tachyons} in its linear approximation.

For the {\bf EGBT} we have
\begin{enumerate}
\item ${d+1\choose 2}$ unknown metric functions,
\item ${d+1\choose 2}$ field equations 2nd-order (but not quasilinear)
\item $d$ differential identities of the 3rd-order,
\item gauge freedom $g_{i^{\prime}j^{\prime}}(x^{\prime}) = {\partial
x^i\over\partial x^{i^{\prime}}}{\partial x^j\over\partial
x^{j^{\prime}}} g_{ij}(x)$.
\end{enumerate}
Repeating reasoning given in the proof of the {\it Proposition 1} we
easily get in the case
\begin{eqnarray}
Z_n(d)&=& {d+1\choose 2}{d\brack n} - d{d\brack d+1}\nonumber \\
&-& \Bigl\{{d+1\choose 2}{d\brack n-2} - d{d\brack n-3}\Bigr\}.
\end{eqnarray}
By use the asymptotic formulas (20)-(21) one can easily obtain from the last formula 
\begin{eqnarray}
Z_n(d)&\asymp& {d\brack n}{d(d-1)d-3)\over n}\nonumber \\
&\asymp& d(d-3){d-1\brack n}.
\end{eqnarray}
We see that in this case 
\begin{equation}
S_E(d) = d(d-1)(d-3) , ~~N_{DF}(d) = d(d-3),
\end{equation}
i.e., we have in the case the same numbers $S_E(d)$ and $N_{DF}(d)$ as
in {\bf GR}.

Thus the {\bf EGBT} has the {\it smallest} numbers $S_E(d)$ and
$N_{DF}(d)$ among the {\bf PMQG}, i.e., this quadratic theory of gravity
has the {\it strongest} field equations among {\bf PMQG}. 

Following Einstein [1] the {\bf EGBT} is the {\it best one theory} from
the all set of the {\bf PMQG} because it has the strongest field
equations.

On the other hand for a standard {\bf PGT}\footnote{With or without
terms quadratic in torsion in its Lagrangian $L_g$.} we have the
following 
\leftline{\it Proposition 2}

\begin{eqnarray}
Z_n(d) &=& {d(d+1)\over 2}{d\brack n} + {d(d-1)\over 2}d{d\brack n-1} -
d{d\brack n+1}\nonumber \\
&-&\Bigl\{d^2{d\brack n-2} + {d(d-1)\over 2}d{d\brack n-3} -
{d(d-1)\over 2}{d\brack n-4} - d{d\brack n-3}\Bigr\}\nonumber \\
&\asymp& {d\brack n}{d(d+1)(d-1)(d-2)\over n}\asymp
d(d+1)(d-2){d-1\brack n},
\end{eqnarray}
i.e., here we have 
\begin{equation}
S_E(d) = d(d+1)(d-1)(d-2), ~~N_{DF}(d) = d(d+1)(d-2).
\end{equation}
Iff $d=4$ then we get from the last equation
\begin {equation}
S_E(4) = 120, ~~N_{DF}(4) = 40.
\end{equation}

The proof of the {\it Proposition 2} is very like to the proof of the
{\it Proposition 1}. Namely, in the formula (35), likely as it was in
the formula (19), the first two terms on the right give the total number
of the nth-order coefficients and the other terms before the sign
$\asymp$ give numbers of independent conditions imposed on these
nth-order coefficients: $d{d\brack n+1}$ follow from gauge freedom and
$\Bigl\{d^2 {d\brack n-2} + d{d\choose 2}{d\brack n-3} - {d\choose
2}{d\brack n-4} - d{d\brack n-3}\Bigr\}$ conditions follow from the
vacuum field equations and from differential identities which are
satisfied by them (see e.g. [6]). By use of the asymptotic formulas
(20)-(21) one can easily obtain the above given expressions on $S_E(d)$
and $N_{DF}(d)$ for a typical {\bf PGT}.

Comparing (23), (24) with (36) we see that a typical 3rd-order {\bf PGT}
has {\it much more greater strength and number dynamical degrees of
freedom} than a typical 4th-order {\bf PMQG}, i.e., a typical {\bf PMQG}
{\it has much more stronger} field equations than a typical {\bf PGT}.

Note also that the formal limes
\begin{equation}
\displaystyle\lim_{d\to\infty}{S_E^{~PGT}(d)\over S_E^{~PMQG}(d)} =
\infty,
\end{equation}
i.e., it is infinite. This means that if $d$ is growing then the vacuum
field equations of a {\bf PMQG} become more and more stronger in
comparison with the vacuum field equations of a typical {\bf PGT}.

We will finish this Section with the following conclusions:
\begin{enumerate}
\item A typical {\bf PMQG} obtained by use {\it Hilbert variational
principle} has much more stronger field equations than a competitive
{\bf PGT} with torsion obtained by use {\it Palatini variational
principle}. 
\item In $d\geq 4$ dimensions the 2nd-order {\bf EGBT} is physically
distinguished among the all set of the metric-compatible and quadratic in
curvature  gravity theories (no tachyons and ghosts, the strongest field
equations). 
\end{enumerate}
\section{Strengths $S_E(d)$ and numbers dynamical degrees of freedom
$N_{DF}(d)$ for other alternative gravity theories in $d\geq 4$
dimensions} 
In order to our paper was not too longer, we give in this Section only
the results of (rather simple) calculations without details.
We will begin with Einstein-Cartan-Sciama-Kibble ({\bf ECSK}) theory of
gravity.
\subsection{{\bf ECSK} theory of gravity}
As it is commonly known that this theory of gravity has the same
linear gravitational Lagrangian $L_g$ as {\bf GR} has, i.e, in this theory
$L_g = L_E = \chi R$ [42,43,44]; but we admit in {\bf ECSK} theory the
metric-compatible connection with nonzero torsion.

One can easily calculate that in this case (see e.g. [6])
\begin{eqnarray}
Z_n(d) &=& {d(d+1)\over 2}{d\brack n} - d{d\brack n+1} + d^2{(d-1)\over
2}{d\brack n-1}\nonumber \\
&-& \Bigl\{d^2{d\brack n-2} + d^2{(d-1)\over 2}{d\brack n-1} -
{d(d-1)\over 2}{d\brack n-2} - d{d\brack n-3}\Bigr\}\nonumber \\
&\asymp& d(d-1)(d-3){d\brack n}\asymp d(d-3){d-1\brack n}.
\end{eqnarray}
We see from the above formula that 
\begin{equation}
S_E(d) = d(d-1)(d-3), ~~N_{DF}(d) = d(d-3)
\end{equation}
in the case.

Iff $d=4$ then we have $S_E(4) = 12 , ~~ N_{DF}(4) = 4$.

So, in the framework of the {\bf ECSK} theory we have the same numbers
$S_E(d)$ and $N_{DF}(d)$ as in {\bf GR}. It is not surprising because we
have here the same vacuum field equations as in {\bf GR}.
\subsection{Teleparallel equivalent of {\bf GR} ({\bf TEGR})}
{\bf TEGR} is a formal rephrasing, step-by-step, the all formalism of
{\bf GR} in terms of the teleparallel Weitzenb\"ock connection
$\Gamma^i_{~kl}$ and its torsion $T^i_{~kl} := \Gamma^i_{~lk} -
\Gamma^i_{~kl}$ (see e.g. [23-25]). The Weitzenb\"ock teleparallel connection
and torsion are determined by a distinguished tetrads (or other
anholonomic frames) field. The gravitational Lagrangian $L_g$ of the
{\bf TEGR} is the Einsteinian Lagrangian of {\bf GR} $L_g = \chi R$ formally rewritten
in terms of Weitzenb\"ock torsion and it is quadratic function of this
torsion. 

In this case we also have the same values of the numbers $S_E(d)$ and
$N_{DF}(d)$ as in {\bf GR} and in {\bf ECSK} theory, i.e., we have 
\begin{equation}
S_E(d) = d(d-1)(d-3), ~~N_{DF}(d) = {S_E(d)\over d-1} = d(d-3).
\end{equation}
Iff $d = 4$ then we obtain from the above expressions $S_E(4) = 12,
~~N_{DF}(4) = 4$. 
\subsection{Teleparallel ``new general relativity'' ({\bf TNGR}) of
Hayashi and Shirafuji}
It is also the theory of gravitation in the Weitzenb\"ock spacetime
[26,27] which is determined by a quadruplet of linearly independent parallel
vector fields $h_{(a)}^{~~~i}$. Gravitational Lagrangian $L_g$ is
quadratic in irreducible parts of teleparallel torsion (determined by
$h_{(a)}^{~~~i}$) with respect to the group of global Lorentz transformations and contains three
free parameters. But this Lagrangian is constructed {\it independently}
of the Einstein Lagrangian $L_E = \chi R$, i.e., {\bf TNGR} {\it is not a
simple refrasing } of {\bf GR}, like {\bf TEGR}. 

One can easily calculate that in this case we have
\begin{eqnarray}
Z_n(d)& =& d^2{d\brack n} - d{d\brack n+1} -\Bigl(d^2{d\brack n-2} -
d{d\brack n-3}\Bigr) \nonumber \\
&\asymp& {d\brack n}{2d(d-1)(d-2)\over n}\asymp 2d(d-2){d-1\brack n}.
\end{eqnarray}
We see that in this case
\begin{equation}
S_E(d)= 2d(d-1)(d-2), ~~ N_{DF}(d) = 2d(d-2).
\end{equation}
For $d=4$ this gives $S_E(4) = 48, ~~N_{DF}(4) = 16$.

One can easily see that in this case we have the same values of the
numbers $S_E(d)$ and $N_{DF}(d)$ as for a typical {\bf PMQG} in general
case. We conclude from the above interesting fact that probably a typical
{\bf PMQG} {\it can be refrased} in terms of the Weitzenb\"ock teleparallel
connection with $L_g$ {\it quartic} in teleparallel torsion (like reformulation of {\bf GR} onto {\bf
TEGR}), i.e., we have some kind of the dynamical equivalence between {\bf PMQG} and
{\bf TNGR}. 
\subsection{Scalar-tensor gravity theories $\equiv$ {\bf GR} + scalar
field} 
The theories of gravity of such a kind follow e.g., from low energy
limit of superstrings [28,29] and contain, as a special case, {\it Jordan-Brans-Dicke}
theory. 

In this case we have (in {\it Einstein frame} or in {\it Jordan frame}) 
\begin{eqnarray}
Z_n(d)&=& {d(d+1)\over 2}{d\brack n} - d{d\brack n+1} + {d\brack
n}\nonumber \\
&-&\Bigl({d(d+1)\over 2}{d\brack n-2} - d{d\brack n-3}\Bigr) - {d\brack
n-2}\nonumber \\
&\asymp&{d\brack n}{(d-1)^2(d-2)\over n}\asymp (d-1)(d-2){d-1\brack n}.
\end{eqnarray}
It follows from the last formula
\begin{equation}
S_E(d) = (d-1)^2(d-2), ~~N_{DF}(d) = (d-1)(d-2).
\end{equation}
Iff $d=4$ then we have $S_E(4) = 18, ~~N_{DF}(4) = 6$.

Note that in these theories of gravity exist two distinguished kind of
frames: {\it Jordan frame} and {\it Einstein frame} which are {\it not
equivalent physically} in general in presence of matter. [45]. These frames are connected by
conformal rescalling of the metric. In our opinion the general physical
non-equivalenece of these two distinguished frames  is a defect of the scalar-tensor theories
of gravity. 
\subsection{The most general {\it metric-affine gauge thery of gravity}
({\bf MAG})}
In {\bf MAG} the spacetime geometry is a metric-affine geometry with the
gravitational field strengths nonmetricity $Dg_{ik}$, torsion
$T^i$ and curvature $\Omega^i_{~k}$ (see e.g. [30--33]). The independent
geometrical variables are: metric $g_{ik}$, tetrad $h_{(a)}^{~~~i}$ and
Lorentz connection $\Gamma^{(a)}_{~~~(b)}{}_i$. The most general parity
conserving quadratic Lagrangian $L_g$ is expressed in terms of the
irreducible pieces of the nonmetricity $Dg_{ik}$, torsion $T^i$ and
curvature $\Omega^i_{~k}$ and contains 28 dimensionless free parameters
(apart from $\chi$, cosmological constant $\Lambda$ and strong coupling
$\rho$). The gravitational field equations are here of 2nd-order with
respect to metric, tetrad and Lorentz connection. The total number of
these equations is equal ${d(d+1)\over 2} +d^2 + d^3$.

One can obtain (by use Lagrange multipliers technics) the all considered in this paper theories of gravity (including
{\bf GR} and {\bf ECSK} theory) as special cases of {\bf MAG}.

It is easy to calculate that in the case of {\bf MAG} we have
\begin{eqnarray}
Z_n(d)& = &  {d\choose 2}{d\brack n} + d^2{d\brack n} + d^3{d\brack n} -
d{d\brack n+1}\nonumber \\
&-&{d\choose 2}{d\brack n-2} - d^2{d\brack n-2} -d^3 {d\brack n-2}
\nonumber \\
&-& d^2{d\brack n-2} + d{d\brack n-3} + d^2{d\brack n-3}\nonumber \\
&\asymp& {d\brack n}{d(d-1)(2d^2 + 2d -3)\over n}\asymp{d-1\brack
n}d(2d^2 +2d -3).
\end{eqnarray}
It follows from the above expression that we have for {\bf MAG} 
\begin{equation}
S_E(d) = d(d-1)(2d^2 + 2d - 3), ~~N_{DF}(d) = d(2d^2 + 2d - 3).
\end{equation}
Iff $d=4$ then we get $S_E(4) = 444, ~~N_{DF}(4) = 148$.

One can see from these results that the vacuum field equations of {\bf
MAG} are very week.
\section{Conclusions}
We would like to finish our paper with the following conclusions: 
\begin{enumerate}
\item The {\bf PMQG} theories obtained by use {\it Hilbert variational
principle} have much more stronger field equations then the competitive
{\bf PGT} obtained by use {\it Palatini variational principle}.

Among the all metric-compatible quadratic gravity in $d\geq 4$ dimensions the 2nd-order {\bf EGBT} has the strongest
field equations, i.e., this theory is the best one quadratic and metric-compatible theory of
gravity. 
\item If we confine to the metric-compatible theories of gravity with
linear (in curvature) gravitational Lagrangian $L_g$ then we will have
the two strongest theories of gravity: {\bf GR} and {\bf ECSK}
theory\footnote{Of course the same strength and number dynamical freedom
have any gravity theory with $L_g = \chi R +$ term quadratic in
torsion whose vacuum equations reduce to the vacuum Einstein
equations.}. It seems to us that the experimental evidence and {\it Ockham
razor} favorize {\bf GR}.
\item Probably any {\bf PMQG} can be reformulated (like Einsteinian {\bf
GR}) into a suitable teleparallel theory of gravity with a gravitational
Lagrangian quartic in torsion.
\item {\bf MAG} has too week field equations and too much free
parameters (28) in order to be a useful model of a quadratic theory of
gravity. 
\end{enumerate}
\centerline{\bf Acknowledgements}
The author would like to thank Prof. Rainer Schimming for suggesting
this research project and for his very useful hints and comments.

The author would like also to thank the {\bf EMA} University in Greifswald
and the Max-Planck-Institut f\"ur Gravitationsphysik in Golm where most of
this work has been done for warm hospitality. Especially, he would like
to thank {\bf DAAD}, Bonn, for financial support.

\end{document}